\documentclass[aps,preprintnumbers,nofootinbib,superscriptaddress,twocolumn]{revtex4-1}
\usepackage[utf8]{inputenc}
\usepackage{amsmath,amssymb}
\usepackage{graphicx}
\begin{document}

\preprint{YITP-20-95, IPMU20-0082} 
\title{Covariant entropy bound beyond general relativity}

\author{Taisuke Matsuda}
\affiliation{Center for Gravitational Physics, Yukawa Institute for Theoretical Physics, Kyoto University, 606-8502, Kyoto, Japan}
\author{Shinji Mukohyama}
\affiliation{Center for Gravitational Physics, Yukawa Institute for Theoretical Physics, Kyoto University, 606-8502, Kyoto, Japan}
\affiliation{Kavli Institute for the Physics and Mathematics of the Universe (WPI), The University of Tokyo, 277-8583, Chiba, Japan}

\date{\today}

\begin{abstract}
We propose a covariant entropy bound in gravitational theories beyond general relativity (GR), using Wald-Jacobson-Myers entropy instead of Bekenstein-Hawking entropy. We first extend the proof of the bound known in $4$-dimensional GR to $D$-dimensional GR, $f(R)$ gravity and canonical scalar-tensor theory. We then consider Einstein-Gauss-Bonnet (EGB) gravity as a more non-trivial example and, under a set of reasonable assumptions, prove the bound in the GR branch of spherically symmetric configurations. As a corollary, it is shown that under the null and dominant energy conditions, the generalized second law holds in the GR branch of spherically symmetric configurations of EGB gravity at the fully nonlinear level. 
\end{abstract}

\maketitle

\section{Introduction}

Holography is a universal property of gravity that is expected to hold in a wide class of theories. It states that all information of a $D$-dimensional gravitational theory can be mapped to a ($D-1$)-dimensional non-gravitational theory. As a precursor of holography, Bekenstein~\cite{Bekenstein:1973ur} proposed that a black hole should have entropy proportional to the area of the horizon, based on earlier works uncovering the non-decreasing property of the horizon area~\cite{Christodoulou:1970wf,Penrose:1971uk,Hawking:1971tu}. Bekenstein further claimed that the sum of the black hole entropy and the entropy of matter outside the horizon should not decrease. This statement is what is known as the generalized second law (GSL). The recognition of thermodynamic properties of black holes was then strengthened by Bardeen, Carter and Hawking~\cite{Bardeen:1973gs}, who formulated the four laws of black hole thermodynamics, and by Hawking~\cite{Hawking:1974sw}, who discovered quantum particle creation by black holes, known as Hawking radiation. In particular, the proportionality coefficient between the black hole entropy and the horizon area was fixed to one quarter in the Planck unit through the temperature of Hawking radiation. For these reasons, the black hole entropy is often called Bekenstein-Hawking entropy.

Bekenstein later~\cite{Bekenstein:1980jp} claimed that the validity of GSL requires a universal upper bound on the entropy-to-energy ratio $S/E$ of a gravitationally stable, thermodynamic system of matter, 
\begin{equation}
 S/E \lesssim 2 \pi R\,,  \label{eqn:Bekensteinbound}
\end{equation}
where $R$ is the area radius of a sphere surrounding the system. The argument leading to this inequality is based on a gedanken experiment in which a matter system is thrown to a black hole and the assumption that GSL holds. The bound (\ref{eqn:Bekensteinbound}) is called Bekenstein bound.

Bekenstein bound is known to hold in many (if not all) weakly gravitating systems. However, the original ``derivation'' of the bound was questioned by Unruh and Wald ~\cite{Unruh:1982ic,Unruh:1983ir}, who argued that the bound is not needed for the validity of GSL if the buoyancy force due to the thermal atmosphere near the black hole horizon is taken into account. (See also \cite{Shimomura:1999xp,Gao:2001ut} for extension to charged and rotating black holes.) Hence, the logic of the ``derivation'' is still open to discussion. Nonetheless, the bound (\ref{eqn:Bekensteinbound}) itself holds for weakly gravitating systems in the nature. Furthermore, Casini~\cite{Casini:2008cr} proved a version of the Bekenstein bound, using the non-negativity of the relative entropy. He took a region $V$ on a Cauchy surface and showed that the von Neumann entropy $S_{V}$ (defined by $S_{V}(\sigma) = -\operatorname{Tr}(\sigma\ln\sigma)$) and the modular Hamiltonian $K$ (defined by $\rho^0_V = e^{-K}/\operatorname{Tr}e^{-K}$)  satisfy the following inequality for an arbitrary quantum system.  
 \begin{equation}
  S_{V}(\rho_V) - S_{V}(\rho^0_V) \leq 
  \operatorname{Tr}(K\rho_V)-\operatorname{Tr}(K\rho^0_V)\,,  \label{eqn:Casinibound}
\end{equation}
where $\rho^0_V \equiv \operatorname{Tr}_{-V}\rho^0$ and $\rho_V \equiv \operatorname{Tr}_{-V}\rho$ are the reduced density matrices for the vacuum $\rho^0$ and for an arbitrary quantum state $\rho$, respectively. Here, $\operatorname{Tr}_{-V}$ represents the partial trace over the states on the complementary set of $V$ on the Cauchy surface. The left hand side of (\ref{eqn:Casinibound}) is the increase of the entanglement entropy due to excitations encoded in the quantum state $\rho$ and thus may be interpreted as $S$ in (\ref{eqn:Bekensteinbound}). On the other hand, the right hand side of (\ref{eqn:Casinibound}) is to be interpreted as $2\pi RE$. For example, if $V$ is chosen to be a half a spatial plane ($z>0$) in a $4$-dimensional Minkowski spacetime then $K = \int dxdy \int _0^{\infty}dz\, 2\pi z\mathcal{H}(0,x,y,z)$, where $(t,x,y,z)$ is the Minkowski coordinate system and $\mathcal{H}(t,x,y,z)$ is the Hamiltonian density. (Here, $2\pi z$ in the integrand is the inverse of the local Rindler temperature so that $\rho^0_V$ is equivalent to a thermal state with the Rindler temperature.) Hence, as far as the excitations encoded in $\rho$ have a finite support along the $+z$ direction, the right hand side of (\ref{eqn:Casinibound}) may be interpreted as $2\pi$ times the product of the energy and the size. In this sense, the inequality (\ref{eqn:Casinibound}) proved by Casini may be interpreted as a version of the Bekenstein bound.

While the Bekenstein bound was originally ``derived'' through a gedanken experiment including a strongly gravitating system, i.e. a black hole, the bound itself does not depend on the gravitational constant and holds in nearly flat spacetimes, irrespectively of the nature of gravity. In strongly curved spacetimes, on the other hand, it is not obvious how to define $E$ and $R$. It is nonetheless interesting to speculate how to extend the bound to curved spacetimes. In general relativity (GR), $2\pi ER$ for a spherically symmetric system is bounded from above by $A/4$ in the Planck unit, where $A$ is the area enclosing the system, and thus the Bekenstein bound would imply 
\begin{equation}
 S \leq \frac{A}{4}\,. \label{eqn:S<Sbh}
\end{equation}
This inequality itself is not well-defined in curved spacetimes: for example, the area $A$ depends on the time slicing and can be made arbitrarily small by taking an almost lightlike slice. The speculated inequality (\ref{eqn:S<Sbh}) is nonetheless suggestive as it would state that the entropy of a system surrounded by the area $A$ would be bounded from above by the entropy of a black hole with the same area $A$, i.e. that the state with maximal entropy is a black hole .This is consistent with the fact that gravitational collapse leads to a formation of a black hole and the expectation that the entropy of a system (with or without a black hole) should not decrease.

The most well-known candidate for a covariant version of (\ref{eqn:S<Sbh}) is the one proposed by Bousso~\cite{Bousso:1999xy}, called the covariant entropy bound or Bousso bound. Bousso bound is formulated in arbitrary curved spacetimes and respects the general covariance. The statement of the conjectured bound is as follows. Let $A(B)$ be the area of a connected ($D-2$)-dimensional spatial surface $B$ in a $D$-dimensional spacetime. There are four null congruences projecting away from $B$. Let $L$ be a null hypersurface bounded by $B$ and generated by one of the four null congruences orthogonal to $B$. Let $S_{L}$ be the entropy passing through $L$.  If the expansion of  the congruence is non-positive at every point on $L$ then $L$ is called a lightsheet. It is conjecture that $S_{L}$ for a lightsheet will not exceed a quarter of $A(B)$ in the Planck unit: 
\begin{align}
 S_{L} \leq \frac{A(B)}{4}\,.\label{eqn:Boussobound}
\end{align}
Bousso conjectured that this inequality holds not only for thermodynamic systems sufficiently smaller than the curvature radius but also for large regions of the spacetime. This entropy bound can be interpreted as a formulation of the holographic principle in general spacetime.

Flanagan, Marolf and Wald~\cite{Flanagan:1999jp} then showed two proofs of Bousso bound, supposing that matter is approximated by fluids, under two different sets of assumptions. They also extended the setup so that the lightsheet can be terminated at another connected ($D-2$)-dimensional spatial surface $B'$, and suggested the stronger bound, 
\begin{align}
 S_{L} \leq \frac{A(B)-A(B')}{4}\,, \label{eqn:Boussoboundstronger}
\end{align}
where $A(B')$ is the area of $B'$. In the present paper we consider this version of Bousso bound.

As explained above, the Bekenstein bound is independent of details of of gravitational theories but holds only for weakly gravitating systems . On the other hand, the Bousso's covariant entropy bound is applicable to strongly gravitating systems but specialized in GR. One might therefore expect that different gravitational theories should have different covariant entropy bounds in principle. Given the necessity of modification of GR at short distances toward the quantum gravity theory and many attempts of infrared modification of GR with the hope to address the mysteries of the universe such as the accelerated expansion, it is important to investigate whether and how far it is possible to extend the covariant entropy bound to different gravitational theories. The purpose of the present paper is to initiate this rather general program, focusing particularly on $f(R)$ gravity, canonical scalar-tensor theory and Lovelock gravity.

The rest of the present paper is organized as follows. In Sec.~\ref{sec:Ddim-GR} we extend the previously known proof of the covariant entropy bound in $4$-dimensional GR to $D$-dimensional GR. We also reconsider and clarify motivations of the key assumptions in the proof. In Sec.~\ref{sec:beyondGR} we propose an extension of the covariant entropy bound to gravitational theories beyond GR. As simple examples, in Sec.~\ref{sec:examples} we prove the generalized covariant entropy bound in $f(R)$ gravity and canonical scalar-tensor theory in a thermodynamic limit, using Wald entropy. In Sec.~\ref{sec:Lovelock} we then consider Lovelock gravity, employing Jacobson-Myers entropy. Under a set of reasonable assumptions, we prove the bound in the thermodynamic limit for the GR branch of spherical symmetric configurations in Einstein-Gauss-Bonnet gravity. As a corollary of the proof, we then show that in the GR branch of Einstein-Gauss-Bonnet gravity under the null and dominant energy conditions the generalized second law holds for spherical symmetric configurations at the fully nonlinear level. Sec.~\ref{sec:summary} is devoted to a summary of the paper and some discussions.

\section{Proof in $D$-dimensional GR}
\label{sec:Ddim-GR}

In $4$-dimensional GR, Bousso bound (\ref{eqn:Boussoboundstronger}) in a thermodynamic limit was proved by Flanagan, Marolf and Wald~\cite{Flanagan:1999jp} under two different sets of assumptions. Strominger and Thompson~\cite{Strominger:2003br} then refined the proof by employing a slightly different set of assumptions. In this section we extend the latter proof to a $D$-dimensional spacetime. 

\subsection{Setup}

In a $D$-dimensional spacetime let $L$ be a lightsheet generated by a non-expanding congruence of null geodesics from a connected ($D-2$)-dimensional spatial surface $B$ and terminated on another connected ($D-2$)-dimensional spatial surface $B'$. We choose the affine parameter $\lambda$ on each null geodesic so that $0\leq \lambda \leq 1$ on $L$, that $\lambda=0$ on $B$ and that $\lambda=1$ on $B'$. Let $x=(x^1, \cdots, x^{D-2})$ be a coordinate system on $B$ and $h(x)$ be the determinant of the induced metric on $B$. We can then promote $(x,\lambda)$ to a coordinate system on $L$ by extending $x$ to everywhere on $L$ so that all points on each null geodesic share the same values of $(x^1, \cdots, x^{D-2})$.

We then define a null vector $k^a(x,\lambda)$ on $L$ as $k^a = \pm (\partial /\partial \lambda)^a$, where the sign $\pm$ is chosen so that $k^a$ is future directed, i.e. $+$ (or $-$, respectively) if $L$ is future (or past) directed. We also define a function $\mathcal{A}(x,\lambda)$ on $L$ as
\begin{equation}
 \mathcal{A}(x,\lambda) \equiv \exp \left[ \int_{0}^{\lambda}d\tilde{\lambda} \theta(x,\tilde{\lambda}) \right]\,, \label{eqn:def-calA}
\end{equation}
where $\theta(x,\lambda)$ is the expansion of the null geodesic congruence, i.e. $\theta \equiv \nabla_a (\partial /\partial \lambda)^a$. The integral of a function $f(x,\lambda)$ on $L$ can then be split into an integral over the interval $0\leq \lambda \leq 1$ and an integral on $B$ as
\begin{equation}
 \int_{L} f = \int_{B}d^{D-2}x \sqrt{h(x)} \int_{0}^{1}d\lambda f(x,\lambda)\mathcal{A}(x,\lambda)\,,
\end{equation}
and the difference between the areas of $B$ and $B'$ is expressed as
\begin{equation}
 A(B)-A(B') = \int_{B}d^{D-2}x \sqrt{h(x)} \left[\mathcal{A}(x,0)-\mathcal{A}(x,1)\right]\,.
\end{equation}

In a thermodynamic limit, the matter system admits an entropy flux $D$-vector $s^a$. We then define a function $s(x,\lambda)$ on $L$ as
\begin{equation}
 s \equiv \left. -k_a s^a\right|_{L}\,,
\end{equation}
so that the entropy passing through $L$ is expressed as
\begin{equation}
 S_{L} = \int_{B}d^{D-2}x \sqrt{h(x)} \int_{0}^{1}d\lambda s(x,\lambda)\mathcal{A}(x,\lambda)\,.
\end{equation}

\subsection{Proof}

Strominger and Thompson~\cite{Strominger:2003br} made the following two assumptions in $4$-dimensions. 
\begin{align}
 \text {(i)} & & & s'(x,\lambda)  \leq 2 \pi \mathcal{T}(x,\lambda)\,, \quad \mathcal{T} \equiv T_{a b}k^{a}k^{b}\,,  \nonumber\\
 \text {(ii)} & & & s(x,0) \leq -\frac{1}{4} \mathcal{A}'(x,0) = -\frac{1}{4} \theta(x,0) \,, \label{eqn:assumptions-GR}
\end{align}
where a prime denotes derivative with respect to $\lambda$ and $T_{a b}$ is the stress energy tensor. Physical meaning and justification of these assumptions will be discussed in subsections \ref{subsec:assumption-i} and \ref{subsec:assumption-ii}. In the following we shall employ the same assumptions and prove the Bousso bound in $D$-dimensions.

By the assumption (i) we have
\begin{eqnarray}
 s(x,\lambda) & = & \int_{0}^{\lambda} d\tilde{\lambda} s'(x,\tilde{\lambda}) + s(x,0)\nonumber\\
   & \leq & 2\pi \int_{0}^{\lambda} d\tilde{\lambda}  \mathcal{T}(x,\tilde{\lambda}) + s(x,0)\,.
\end{eqnarray}
The Raychaudhuri equation combined with the Einstein equation implies that
\begin{equation}
 \mathcal{T} = -\frac{D-2}{8\pi}\frac{G''}{G} - \frac{1}{8\pi}\sigma_{ab}\sigma^{ab} 
  \leq -\frac{D-2}{8\pi}\frac{G''}{G}\,,
\end{equation}
where $\sigma_{ab}$ is the shear of the null geodesic congruence and we have defined
\begin{equation}
 G(x,\lambda) \equiv \left[ \mathcal{A}(x,\lambda)\right]^{\frac{1}{D-2}}\,. 
\end{equation}
Hence we have
\begin{eqnarray}
 s(x,\lambda) & \leq & 2\pi \int_{0}^{\lambda} d\tilde{\lambda}  
  \left[ -\frac{D-2}{8\pi}\frac{G''(x,\bar{\lambda})}{G(x,\bar{\lambda})}  \right] + s(x,0)\nonumber\\
& = & \frac{D-2}{4}\left[\frac{G'(x,0)}{G(x,0)}-\frac{G'(x,\lambda)}{G(x,\lambda)}\right] \nonumber\\
& &  - \frac{D-2}{4}\int_{0}^{\lambda}d\tilde{\lambda}  \left[\frac{G'(x,\tilde{\lambda})}{G(x,\tilde{\lambda})}\right]^2 + s(x,0)\,.
\end{eqnarray}
By using the assumption (ii) and $G(x,0) = \mathcal{A}(x,0) = 1$, we thus have
\begin{eqnarray}
 s(x,\lambda) & \leq & -\frac{D-2}{4}\frac{G'(x,\lambda)}{G(x,\lambda)} - \frac{D-2}{4}\int_{0}^{\lambda}d\tilde{\lambda}  \left[\frac{G'(x,\tilde{\lambda})}{G(x,\tilde{\lambda})}\right]^2 \nonumber\\
& \leq & -\frac{D-2}{4}\frac{G'(x,\lambda)}{G(x,\lambda)} = -\frac{1}{4}\frac{\mathcal{A}'(x,\lambda)}{\mathcal{A}(x,\lambda)} \,.
\end{eqnarray}
Integrating this inequality over $L$, we obtain the Bousso bound (\ref{eqn:Boussoboundstronger}).

\subsection{Motivation of assumption (i)}
\label{subsec:assumption-i}

As already argued by Strominger and Thompson~\cite{Strominger:2003br}, the assumption (i) in (\ref{eqn:assumptions-GR}) simply requires that the rate of change of the entropy flux is less than the energy flux. For this reason, it is expected that the assumption (i) should hold as far as the thermodynamic approximation is valid.

The assumption (i) can also be considered as a consequence of a version of Bekenstein bound. Let us consider a subspace $\Delta B$ of $B$ and an interval $[\lambda_1, \lambda_2]$ ($\subset [0,1]$) that are sufficiently smaller than the curvature radius. Suppose that a version of Bekenstein bound of the following form holds for the subspace $\Delta L \equiv \{ (x,\lambda) |\  x \in \Delta B\,, \  \lambda \in [\lambda_1,\lambda_2] \}$ of $L$. 
\begin{eqnarray}
& & \int_{\Delta B}d^{D-2}x\sqrt{h(x)}\int_{\lambda_1}^{\lambda_2}d\lambda \mathcal{A}(x,\lambda)
  \nonumber\\
 & & \times \left[ s(x,\lambda)-s(x,\lambda_1) - 2\pi (\lambda-\lambda_1)\mathcal{T}(x,\lambda) \right] \leq 0\,,
\end{eqnarray}
which is similar~\footnote{Here, we consider the difference $s(x,\lambda)-s(x,\lambda_1)$ instead of $s(x,\lambda)$ itself. Otherwise, the bound in the limit $\lambda-\lambda_1 \to 0$ would require $s(x,\lambda_2)\leq 0$. In Casini's proof, on the other hand, the difference between the state of interest and the vacuum is considered.} to the version of Bekenstein bound proved by Casini (see (\ref{eqn:Casinibound}) and discussion after that). This holds for various choices of $\Delta B$ and $[\lambda_1,\lambda_2]$ if and only if 
\begin{equation}
 s(x,\lambda) - s(x,\lambda_1) \leq 2\pi (\lambda-\lambda_1) \mathcal{T}(x,\lambda)\,. 
 \end{equation}
From this, we obtain the assumption (i) in the limit $\lambda-\lambda_1\to 0$.

\subsection{Motivation of assumption (ii)}
\label{subsec:assumption-ii}

The assumption (ii) in (\ref{eqn:assumptions-GR}) is just to prevent Bousso bound (\ref{eqn:Boussoboundstronger}) from being violated infinitesimally at the beginning of $L$ and thus is mandatory.

Moreover, if we consider $B$ in a sufficiently flat region of spacetime then the assumption (ii) follows from Bekenstein bound. For simplicity let us consider an approximately Minkowski spacetime region with the radial coordinate $r$ and the time coordinate $t$, and suppose that $B$ is a sphere of radius $r=r_0$ at time $t=t_0$. Let us then consider a future-directed lightsheet $L$ and another connected ($D-2$)-dimensional spatial surface $\tilde{B}$ on $L$ at $\lambda=\delta\lambda$, $r=\tilde{r}_0$ and $t=\tilde{t}_0$. While $L$ is by definition generated by a congruence of future-directed ingoing null geodesics, we also consider the congruence of future-directed outgoing null geodesics from $B$ and call the intersection of $t=\tilde{t}_0$ and the latter congruence $\bar{B}$. Obviously, the radius of $\bar{B}$ is $\bar{r}_0=r_0+\delta r$, where $\delta r = r_0-\tilde{r}_0$ ($>0$). We call the region between $\tilde{B}$ and $\bar{B}$ on the $t=\tilde{t}_0$ hypersurface $\delta\Sigma$. The entropy passing through the portion $\delta L$ of $L$ between $B$ and $\tilde{B}$ is 
\begin{equation}
 \delta S = \int_B d^{D-2}x\sqrt{h(x)}\int_{0}^{\delta\lambda}d\lambda s(x,\lambda)\mathcal{A}(x,\lambda)\,. \label{eqn:deltaS}
\end{equation}
Assuming the dominant energy condition, the entropy flux passing through the portion $\delta L$ inevitably passes $\delta\Sigma$ and thus the second law of thermodynamics states that
\begin{equation}
 \delta S \leq \delta \tilde{S}\,, 
\end{equation}
where $\delta\tilde{S}$ is the entropy contained in $\delta\Sigma$. We now divide $\delta \Sigma$ into small pieces with sizes of order $2\delta r$ (the distance between $\tilde{B}$ and $\bar{B}$), apply Bekenstein bound to each piece, and then consider the total sum. Demanding that the total energy in $\delta\Sigma$ be not exceeding the mass $(D-2)\mathrm{a}_{D-2}\bar{r}_0^{D-3}/(16\pi)$ of a spherically symmetric black hole with the area radius $\bar{r}_0$, where $\mathrm{a}_{D-2}= (D-1)\pi^{\frac{D-1}{2}}/\Gamma((D+1)/2)$ is the surface area of the unit $(D-2)$-sphere, we obtain 
\begin{eqnarray}
 \delta \tilde{S} & \leq & 2\pi \cdot 2\delta r \cdot \frac{D-2}{16\pi}\mathrm{a}_{D-2}r_0^{D-3} + \mathcal{O}(\delta r^2) \label{eqn:deltatildeS}\\
& = & \frac{1}{4}\int_Bd^{D-2}x\sqrt{h(x)}\left[ \mathcal{A}(x,0)-\mathcal{A}(x,\delta\lambda)\right] + \mathcal{O}(\delta r^2)\,.\nonumber
\end{eqnarray}
Combining (\ref{eqn:deltaS})-(\ref{eqn:deltatildeS}) and taking the limit $\delta\lambda \to 0$, we obtain the assumption (ii).

For a past-directed lightsheet $L$, we consider the congruence of future-directed in-going null geodesics from $\tilde{B}$, consider $\bar{B}$ as the intersection of such a congruence with the $t=t_0$ hypersurface and $\delta \Sigma$ as the region between $B$ and $\bar{B}$ on the $t=t_0$ hypersurface. Then, from Bekenstein bound applied to pieces of $\delta\Sigma$ and the second law of thermodynamics, we again deduce the assumption (ii).

\section{Extension to gravity beyond GR}
\label{sec:beyondGR}

GR is a neither unique nor complete theory of gravity but an effective theory valid in a certain range of scales. At extremely short or/and long distances, GR may break down and extra degrees of freedom or/and additional terms in the gravity action may kick in. On the other hand, holography is expected  to be a universal property of gravity that hold in various theories. It is therefore natural to ask whether the covariant entropy bound can be extended to theories of gravity beyond GR.

As explained in introduction, the covariant entropy bound (\ref{eqn:Boussobound}) or its refined version (\ref{eqn:Boussoboundstronger}) is a covariant and precise version of the statement that the entropy of a system surrounded by an area $A$ should be bounded from above by the entropy of a black hole with the same area $A$. Indeed, the right hand side of (\ref{eqn:Boussobound}) is the black hole entropy in GR and that of (\ref{eqn:Boussoboundstronger}) is difference between the black hole entropy corresponding to the final and initial states. In gravitational theories beyond GR in $D$-dimensions, if there is a formula of black hole entropy that can be formally applied to not only black hole horizons but also other connected ($D-2$)-dimensional spatial surfaces, it is rather natural to conjecture that the entropy $S_{L}$ passing through a lightsheet $L$, suitably generalized as explained below, should satisfy
\begin{equation}
 S_{L} \leq S_{\mathrm{bh}}(B) - S_{\mathrm{bh}}(B')\,, \label{eqn:Boussoboundgeneralized}
\end{equation}
where $B$ is the connected ($D-2$)-dimensional spatial surface from which $L$ is generated, $B'$ is another connected ($D-2$)-dimensional spatial surface at which $L$ is terminated, and $S_{\mathrm{bh}}(B)$ (and $S_{\mathrm{bh}}(B')$) is the formula of black hole entropy applied to $B$ (and $B'$, respectively). We call this inequality a {\it generalized covariant entropy bound} for a generalized lightsheet $L$ defined below.

Since the left hand side of (\ref{eqn:Boussoboundgeneralized}) is non-negative,  a generalized lightsheet should be defined so that the right hand side is always non-negative. This suggests the following definition of a generalized lightsheet in gravitational theories beyond GR. Let us consider a congruence of null geodesics from a connected $(D-2)$-dimensional spatial surface $B$ and terminated at another connected $(D-2)$-dimensional spatial surface $B'$. We assume that there is no singularity or caustics of the congruence of null geodesics between $B$ and $B'$. Again, without loss of generality, we normalize the affine parameter $\lambda$ on each geodesic so that $\lambda=0$ on $B$ and that $\lambda=1$ on $B'$. We then introduce a coordinate system $x=(x^1,\cdots,x^{D-2})$ on $B$ and promote $(x,\lambda)$ to a coordinate system on the null hypersurface between $B$ and $B'$ by extending $x$ along each geodesic so that all points on each null geodesic share the same values of $(x^1,\cdots,x^{D-2})$. This naturally defines a one-parameter family of connected $(D-2)$-dimensional spatial surfaces $B(\lambda)$ parameterized by $\lambda$ (with $B(0)=B$ and $B(1)=B'$) and $x$ can be considered as a coordinate system on each $B(\lambda)$. Suppose that there is a well-defined formula of black hole entropy that can be formally applied to $B(\lambda)$ ($0\leq \lambda \leq 1$) and that it is of the form
\begin{equation}
 S_{\mathrm{bh}}(B(\lambda)) = \int_{B(\lambda)} d^{D-2}x\sqrt{h(x,\lambda)}s_{\mathrm{bh}}(x,\lambda)\,,
  \label{eqn:SBH-general}
\end{equation}
where $h(x,\lambda)$ is the determinant of the induced metric on $B(\lambda)$ and $s_{\mathrm{bh}}(x,\lambda)$ is a function on $B(\lambda)$. (In GR, $s_{\mathrm{bh}}(x,\lambda)=1/4$.) We then define a {\it generalized expansion} $\Theta(x,\lambda)$ as
\begin{equation}
 \Theta(x,\lambda) \equiv \frac{\partial}{\partial \lambda} \ln \left[ \sqrt{h(x,\lambda)}s_{\mathrm{bh}}(x,\lambda) \right]\,. \label{eqn:expansiongeneralized}
\end{equation}
If $\Theta(x,\lambda)\leq 0$ for ${}^{\forall}x$ and $0\leq {}^{\forall}\lambda \leq 1$ then we call the null hypersurface between $B$ and $B'$ generated by the congruence of null geodesics a {\it generalized lightsheet}.

For later convenience, we define a null vector $k^a(x,\lambda)$ on a generalized lightsheet $L$ as $k^a = \pm (\partial /\partial \lambda)^a$, where the sign $\pm$ is chosen so that $k^a$ is future directed, i.e. $+$ (or $-$, respectively) if $L$ is future (or past) directed.

Therefore, for a given formula of black hole entropy in a gravitational theory of interest, if the formula is applicable to any connected $(D-2)$-dimensional spatial surfaces then one can define the generalized lightsheet in terms of the generalized expansion (\ref{eqn:expansiongeneralized}) and then formulate the generalized covariant entropy bound (\ref{eqn:Boussoboundgeneralized}). This is a rather general program. However, a well-defined formula of black hole entropy that can be formally applied to connected $(D-2)$-dimensional spatial surfaces away from a black hole horizon is not always available for gravitational theories beyond GR. In the present paper, we shall only consider $f(R)$ gravity, canonical scalar-tensor theory and Lovelock gravity. We expect that there should be other theories of gravity for which the generalized covariant entropy bound can be formulated, but exploring such possibilities is beyond the scope of the present paper and we shall focus on the three types of theories only.

\section{Simple examples}
\label{sec:examples}

In this section we consider $f(R)$ gravity and a canonical scalar-tensor theory as simple examples to illustrate the idea of the generalized covariant entropy bound proposed in the previous section.

\subsection{Wald entropy}

As proposed in the previous section, in order to formulate the generalized covariant entropy bound in a gravitational theory beyond GR, we need to specify a definition of black hole entropy. In this section we shall adopt Wald entropy which is defined as a Noether charge~\cite{Wald:1993nt} and written as~\cite{Iyer:1994ys}
\begin{equation}
 S_{\mathrm{bh}}(\mathcal{H})  = -2\pi \int_{\mathcal{H}} d^{D-2}x \sqrt{h} E_{R}^{abcd}\epsilon_{ab}\epsilon_{cd}\,, \label{eqn:Waldentropy}
\end{equation}
where $\mathcal{H}$ is the bifurcation surface of a Killing horizon, $h$ is the determinant of the induced metric on $\mathcal{H}$, $E_{R}^{abcd}$ is the functional derivative of the gravitational action with respect to the Riemann tensor $R_{abcd}$ with the metric and the connection held fixed, and $\epsilon_{ab}$ is the binormal to $\mathcal{H}$. In this section we apply this formula not only to $\mathcal{H}$ but also to the family of connected spatial $(D-2)$ surfaces $B(\lambda)$ introduced in the previous section. This corresponds to 
\begin{equation}
 s_{\mathrm{bh}} = -2\pi E_{R}^{abcd}\epsilon_{ab}\epsilon_{cd}\,. 
\end{equation}
In this section we shall adopt this choice of $s_{\mathrm{bh}}$ for $f(R)$ gravity and a canonical scalar-tensor theory.

\subsection{f(R) gravity}

For $f(R)$ gravity described by the action
\begin{equation}
 I_{f} =\frac{1}{16\pi} \int d^Dx \sqrt{-g} f(R)\,, \label{eqn:action-f(R)}
\end{equation}
we have
\begin{equation}
 s_{\mathrm{bh}} = \frac{1}{4}f'(R)\,. \label{eqn:sBH-f(R)}
\end{equation}
Throughout this section we assume that 
\begin{equation}
f'(R)>0\,, 
\end{equation}
in order to ensure that tensorial gravitational waves have a kinetic term with a positive coefficient. The generalized expansion is then given by (\ref{eqn:expansiongeneralized}) with (\ref{eqn:sBH-f(R)}). 

We shall rewrite the generalized covariant entropy bound (\ref{eqn:Boussoboundgeneralized}) with (\ref{eqn:SBH-general}) and (\ref{eqn:sBH-f(R)}) in a way that can be easily proved. For this purpose, we note that the action (\ref{eqn:action-f(R)}) is equivalent to 
\begin{equation}
 \tilde{I}_{f} =\frac{1}{16\pi} \int d^Dx \sqrt{-g} \left[ f'(\varphi)R + f(\varphi) - \varphi f'(\varphi) \right]\,, \label{eqn:equivalent-action-f(R)}
\end{equation}
provided that $f''(R)\ne 0$~\footnote{If $f''(R)=0$ then $f(R)$ is linear in $R$ and the theory is reduced to GR.}. On introducing the metric in Einstein frame $g^{\mathrm{E}}_{ab}$ and redefining the scalar field as
\begin{equation}
 g^{\mathrm{E}}_{ab} \equiv [f'(\varphi)]^{\frac{2}{D-2}} g_{ab}\,, \quad
  \phi \equiv \sqrt{\frac{2(D-1)}{D-2}}\ln f'(\varphi)\,, \label{eqn:def-Einsteinframe-f(R)}
\end{equation}
the equivalent action (\ref{eqn:equivalent-action-f(R)}) is rewritten as
\begin{equation}
 \tilde{I}_{f} = \int d^Dx\sqrt{-g^{\mathrm{E}}}\left[ \frac{R^{\mathrm{E}}}{16\pi} - \frac{1}{2}g_{\mathrm{E}}^{ab}\partial_a\phi\partial_b\phi - V(\phi)\right]\,, \label{eqn:action-f(R)-Einsteinframe}
\end{equation}
where 
\begin{equation}
 V(\phi) \equiv  \frac{1} {[f'(\varphi)]^{\frac{D}{D-2}}}\left(\varphi f'(\varphi)-f(\varphi)\right)\,.
\end{equation}
The equivalent action (\ref{eqn:action-f(R)-Einsteinframe}) written in terms of $g^{\mathrm{E}}_{\mu\nu}$ is nothing but the Einstein-Hilbert action coupled to the canonical scalar field $\phi$ with the potential $V(\phi)$. Therefore black hole entropy in Einstein frame is given by Bekenstein-Hawking formula as
\begin{equation}
 S_{\mathrm{bh}}^{\mathrm{E}}(B(\lambda)) = \frac{1}{4}\int_{B(\lambda)}d^{D-2}x \sqrt{h^{\mathrm{E}}(x,\lambda)}\,, \label{eqn:SBH-Einsteinframe}
\end{equation}
where $h^{\mathrm{E}}(x,\lambda)$ is the determinant of the induced metric on $B(\lambda)$ in Einstein frame.

Null geodesics are mapped to null geodesics by the conformal transformation. However, the affine parameter $\lambda$ in the original frame is not an affine parameter in Einstein frame in general. We thus introduce the affine parameter $\lambda^{\mathrm{E}}$ in Einstein frame as
\begin{equation}
 \lambda^{\mathrm{E}} = \frac{1}{\mathcal{N}(x)}\int_{0}^{\lambda} [f_{R}(x,\lambda')]^{\frac{2}{D-2}} d\lambda'\,,
\end{equation}
where 
\begin{equation}
 \mathcal{N}(x) = \int_{0}^{1} [f_{R}(x,\lambda')]^{\frac{2}{D-2}} d\lambda'\,,
\end{equation}
and $f_{R}(x,\lambda)$ is the value of $f'(\varphi) = f'(R)$ at $(x, \lambda)$ on $L$, so that $0\leq \lambda^{\mathrm{E}}\leq 1$ on $L$, that $\lambda^{\mathrm{E}}=0$ on $B$ and that $\lambda^{\mathrm{E}}=1$ on $B'$. Correspondingly, we introduce the null generator $k_{\mathrm{E}}^a$ of $L$ in Einstein frame as
\begin{equation}
 k_{\mathrm{E}}^a \equiv \mathcal{N}(x) [f_{R}(x,\lambda)]^{-\frac{2}{D-2}} k^a\,.
\end{equation}

From the relation between $g^{\mathrm{E}}_{ab}$ and $g_{ab}$ in (\ref{eqn:def-Einsteinframe-f(R)}), it is obvious that $h^{\mathrm{E}}(x,\lambda) = h(x,\lambda)\cdot [f_{R}(x,\lambda)]^2$. Since the equation of motion for $\varphi$ implies $\varphi = R$, Bekenstein-Hawking entropy in Einstein frame (\ref{eqn:SBH-Einsteinframe}) agrees with Wald entropy in the original frame, i.e. (\ref{eqn:SBH-general}) with (\ref{eqn:sBH-f(R)}). Also, the expansion in Einstein frame $\theta_{\mathrm{E}}$ is related to the generalized expansion in the original frame, (\ref{eqn:expansiongeneralized}) with (\ref{eqn:sBH-f(R)}), as
\begin{equation}
 \theta_{\mathrm{E}} = \frac{\partial}{\partial \lambda^{\mathrm{E}}} \ln \sqrt{h^{\mathrm{E}}}
  = \mathcal{N}(x) [f_{R}(x,\lambda)]^{-\frac{2}{D-2}} \Theta\,.
\end{equation}
Hence, $\theta_{\mathrm{E}}$ and $\Theta$ have the same sign, meaning that the definition of a lightsheet in Einstein frame agrees with that of a generalized lightsheet in the original frame.

Therefore, the generalized covariant entropy bound (\ref{eqn:Boussoboundgeneralized}) with (\ref{eqn:SBH-general}) and (\ref{eqn:sBH-f(R)}) holds if quantities in Einstein frame satisfy the assumptions (i) and (ii) in (\ref{eqn:assumptions-GR}), i.e. 
\begin{align}
 \text{(i-f)} & & & \frac{\partial s_{\mathrm{E}}}{\partial \lambda^{\mathrm{E}}}(x,\lambda^{\mathrm{E}})  \leq 2 \pi \mathcal{T}^{\mathrm{E}}(x,\lambda^{\mathrm{E}})\,, \nonumber\\
 \text{(ii-f)} & & & s_{\mathrm{E}}(x,0) \leq -\frac{1}{4} \theta_{\mathrm{E}}(x,0)\,, \label{eqn:assumptions-f(R)-Einsteinframe}
\end{align}
where $s_{\mathrm{E}}$ is the entropy density in Einstein frame, $\mathcal{T}^{\mathrm{E}} \equiv (T^{\mathrm{E}}_{a b}+\partial_{a}\phi\partial_{b}\phi)k_{\mathrm{E}}^{a}k_{\mathrm{E}}^{b}$, and $T^{\mathrm{E}}_{a b}$ is the stress energy tensor of matter in Einstein frame. It is straightforward to rewrite $\text{(i-f)}$ and $\text{(ii-f)}$ in (\ref{eqn:assumptions-f(R)-Einsteinframe}) in terms of quantities in the original frame. Since the entropy $S_{L}$ is invariant under frame transformations, we have $\sqrt{h^{\mathrm{E}}}s_{\mathrm{E}}d\lambda^{\mathrm{E}}=\sqrt{h}sd\lambda$ and thus $s=s_{\mathrm{E}} [f_{R}(x,\lambda)]^{\frac{D}{D-2}}/\mathcal{N}(x)$. The definition of stress energy tensors and the relation between $g^{\mathrm{E}}_{ab}$ and $g_{ab}$ in (\ref{eqn:def-Einsteinframe-f(R)}) imply that $T_{ab} = T^{\mathrm{E}}_{ab} f'(R)$, where $T_{ab}$ is the stress energy tensor of matter in the original frame. Therefore, the condition $\text{(i-f)}$ in (\ref{eqn:assumptions-f(R)-Einsteinframe}) is rewritten as
\begin{align}
 \text{(i-f)} & & & \frac{\partial s}{\partial \lambda}(x,\lambda)  \leq 2\pi \mathcal{T}(x,\lambda) + \Delta_{f}(x,\lambda)\,,
\end{align}
where $\mathcal{T}\equiv T_{ab}k^ak^b$ and
\begin{equation}
 \Delta_{f} \equiv 2\pi f_{R}(x,\lambda)(k^a\partial_a\phi)^2 +\frac{D}{D-2} s\frac{\partial}{\partial\lambda}\ln f_{R}(x,\lambda)\,.
\end{equation}
The first term in $\Delta_{f}$ is always non-negative. The second term in $\Delta_{f}$ is much smaller than $2\pi\mathcal{T}(x,\lambda)$ as far as the rate of change of $\ln f_{R}(x,\lambda)$ is much less than the local temperature of the system as required by the local equilibrium. In this case, the condition $\text{(i-f)}$ holds if quantities in the original frame satisfy the condition (i) in (\ref{eqn:assumptions-GR}), i.e. if
\begin{align}
 \text{(i-f)}' & & & \frac{\partial s}{\partial \lambda}(x,\lambda)  \leq 2\pi \mathcal{T}(x,\lambda)\,.
\label{eqn:assumptions-f(R)-originalframe-i}
\end{align}
On the other hand, the condition $\text{(ii-f)}$ in (\ref{eqn:assumptions-f(R)-Einsteinframe}) is rewritten as
\begin{align}
 \text{(ii-f)} & & & s(x,0) \leq - s_{\mathrm{bh}}(x,0)\Theta(x,0)\,.
  \label{eqn:assumptions-f(R)-originalframe-ii}
\end{align}
By applying the argument in subsection \ref{subsec:assumption-ii} to the system in Einstein frame, one can motivate the condition $\text{(ii-f)}$ in the form of (\ref{eqn:assumptions-f(R)-Einsteinframe}). One can also extend the argument in subsection \ref{subsec:assumption-ii} so that it can be directly applied to the system in the original frame for constant or slowly varying $f'(R)$ to motivate the condition $\text{(ii-f)}$ in the form of (\ref{eqn:assumptions-f(R)-originalframe-ii}) by noting that the effective gravitational constant in this case is inversely proportional to $f'(R)$ and thus that the upper bound on the energy inside a sphere of a given radius is proportional to $f'(R)$.

\subsection{Canonical scalar-tensor theory}

The action of a canonical scalar-tensor theory is
\begin{align}
I_{\rm st} = \frac{1}{16\pi} \int d^Dx \sqrt{-g}\left[ F(\varphi)R - \frac{1}{2}g^{\mu\nu}\partial_{\mu}\varphi\partial_{\nu}\varphi - U(\varphi) \right]\,, \label{eqn:action-ST} 
\end{align}
for which we have 
\begin{align}
s_{\mathrm{bh}} = \frac{1}{4} F(\varphi)\,. \label{eqn:sBH-ST}
\end{align}
Hereafter we assume $F(\varphi) \geq 0$ to avoid tensor ghosts. The  generalized expansion is then given by (\ref{eqn:expansiongeneralized}) with (\ref{eqn:sBH-ST}), and the generalized covariant entropy bound is given by (\ref{eqn:Boussoboundgeneralized}) with (\ref{eqn:SBH-general}) and (\ref{eqn:sBH-ST}).

On introducing the metric in Einstein flame and redefining the scalar field as
\begin{align}
g^{E}_{ab} & = [F(\varphi)]^{\frac{2}{D-2}}g_{ab}\,, \nonumber\\
 \phi & = \int d\tilde{\varphi} \sqrt{F(\tilde{\varphi})^{-1}+\frac{2(D-1)}{D-2}\left[\frac{F^{\prime}(\tilde{\varphi})}{F(\tilde{\varphi})}\right]^2}\,,
\end{align}
we can rewrite the action as
 \begin{align}
I_{\rm st} = \frac{1}{16\pi} \int d^D x \sqrt{-\tilde{g}} \left[\tilde{R} - \frac{1}{2}\tilde{g}^{\mu\nu}\partial_{\mu}\phi\partial_{\nu}\phi - V(\phi)  \right]\,, \label{eqn:action-ST-Einsteinframe}
\end{align}
where 
\begin{equation}
 V(\phi) = [F(\varphi)]^{-\frac{D}{D-2}}U(\varphi)\,.
\end{equation}
The action (\ref{eqn:action-ST-Einsteinframe}) is the Einstein-Hilbert action coupled to a canonical scalar field and thus the black hole entropy in this frame is given by Bekenstein- Hawking formula as 
\begin{align}
 S_{\mathrm{bh}}^{\mathrm{E}}(B(\lambda)) = \frac{1}{4}\int_{B(\lambda)}d^{D-2}x \sqrt{h^{\mathrm{E}}(x,\lambda)}\,,
  \end{align}
where $h^{\mathrm{E}}$ is the determinant of the induced metric in Einstein frame. In the same way as in the case of $f(R)$ gravity, we introduce the affine parameter $\lambda^{\mathrm{E}}$ in Einstein flame as
\begin{equation}
 \lambda^{\mathrm{E}} = \frac{1}{\mathcal{N}(x)}\int_{0}^{\lambda} [F(x,\lambda')]^{\frac{2}{D-2}} d\lambda'\,,
\end{equation}
where 
\begin{equation}
 \mathcal{N}(x) = \int_{0}^{1} [F(x,\lambda')]^{\frac{2}{D-2}} d\lambda'\,,
\end{equation}
and $F(x,\lambda)$ is the value of $F(\varphi)$ at $(x,\lambda)$ on $L$, so that $0\leq \lambda^{\mathrm{E}}\leq 1$ on $L$, that $\lambda^{\mathrm{E}}=0$ on $B$ and that $\lambda^{\mathrm{E}}=1$ on $B'$. Correspondingly, we introduce the null generator $k_{\mathrm{E}}^a$ of $L$ in Einstein frame as
\begin{equation}
 k_{\mathrm{E}}^a \equiv \mathcal{N}(x) [F(x,\lambda)]^{-\frac{2}{D-2}} k^a\,.
\end{equation}

Since $h^E =  h\cdot [F(x,\lambda)]^2$, Bekenstein-Hawking entropy in Einstein flame is equal to Wald entropy in the original flame.  Also, the expansion $\theta_{\mathrm{E}}$ in Einstein flame is related to the generalized expansion $\Theta$ in the original flame as
\begin{equation}
 \theta_{\mathrm{E}} = \frac{\partial}{\partial \lambda^{\mathrm{E}}} \ln \sqrt{h^{\mathrm{E}}}
  = \mathcal{N}(x) [F(\varphi)]^{-\frac{2}{D-2}} \Theta\,,
\end{equation}
and thus $\theta_{\mathrm{E}}$ and $\Theta$ have the same sign. Therefore the definition of a generalized lightsheet in the original flame coincides with that of a lightsheet in Einstein flame.

Thus, the generalized covariant entropy bound holds if quantities in Einstein flame satisfy the assumption (i) and (ii) in (\ref{eqn:assumptions-GR}), i.e.
\begin{align}
\text{(i-s)}  & & & \frac{\partial s_{\mathrm{E}}}{\partial \lambda^{\mathrm{E}}}(x,\lambda^{\mathrm{E}})  \leq 2 \pi \mathcal{T}^{\mathrm{E}}(x,\lambda^{\mathrm{E}})\,, \nonumber\\
\text{(ii-s)} & & & s_{\mathrm{E}}(x,0) \leq -\frac{1}{4} \theta_{\mathrm{E}}(x,0)\,. \label{eqn:assumptions-ST-Einsteinframe}
\end{align}
Following the same logic as in the previous subsection for $f(R)$ gravity, we obtain $s=s_{\mathrm{E}} [F(x,\lambda)]^{\frac{D}{D-2}}/\mathcal{N}(x)$ and $T_{ab} = T^{\mathrm{E}}_{ab} f'(R)$. Therefore, the condition $\text{(i-s)}$ in (\ref{eqn:assumptions-ST-Einsteinframe}) is rewritten as
\begin{align}
\text{(i-s)} & & &  \frac{\partial s}{\partial \lambda}(x,\lambda)  \leq 2\pi \mathcal{T}(x,\lambda) + \Delta_{s}(x,\lambda)\,,
\end{align}
where $\mathcal{T}\equiv T_{ab}k^ak^b$ and
\begin{align}
 \Delta_{s} \equiv 2\pi F(x,\lambda)(k^a\partial_a\phi)^2 +\frac{D}{D-2} s\frac{\partial}{\partial\lambda}\ln F(x,\lambda)\,.
 \end{align}
The first term in $\Delta_{s}$ is always non-negative due to positivity of $F(\varphi)$. The second term in $\Delta_{s}$ is much smaller than $2\pi\mathcal{T}(x,\lambda)$, as far as the fractional change in $F(\varphi)$ is sufficiently small over a distance of order the inverse temperature, as required by local equilibrium. In this case, the condition $\text{(i-s)}$ holds if quantities in the original frame satisfy the condition (i) in (\ref{eqn:assumptions-GR}), i.e. if
\begin{align}
 \text{(i-s)}' & & & \frac{\partial s}{\partial \lambda}(x,\lambda)  \leq 2\pi \mathcal{T}(x,\lambda)\,. \label{eqn:assumptions-ST-originalframe-i}
\end{align}
The condition $\text{(ii-s)}$ in (\ref{eqn:assumptions-ST-Einsteinframe}) is rewritten as
\begin{align}
 \text{(ii-s)} & & & s(x,0) \leq - s_{\mathrm{bh}}(x,0)\Theta(x,0)\,.
 \label{eqn:assumptions-ST-originalframe-ii}
\end{align}
By applying the argument in subsection \ref{subsec:assumption-ii} to the system in Einstein frame, one can motivate the condition $\text{(ii-s)}$ in the form of (\ref{eqn:assumptions-ST-Einsteinframe}). One can also extend the argument in subsection \ref{subsec:assumption-ii} so that it can be directly applied to the system in the original frame for constant or slowly varying $F(\varphi)$ to motivate the condition $\text{(ii-s)}$ in the form of (\ref{eqn:assumptions-ST-originalframe-ii}) by noting that the effective gravitational constant in this case is inversely proportional to $F(\varphi)$ and thus that the upper bound on the energy inside a sphere of a given radius is proportional to $F(\varphi)$.

\section{Lovelock gravity}
\label{sec:Lovelock}

In this section we consider Lovelock gravity in $D$ dimensions described by the action 
\begin{align}
 I & = \frac{1}{16\pi}\int d^Dx\sqrt{-g}\sum_{n=1}^{[(D-1)/2]} \alpha_{n} \mathcal{L}_{n}[g_{\mu\nu}]\,, \nonumber \\
\mathcal{L}_{n}[g_{\mu\nu}] &=\frac{1}{2^{n}} \delta_{\alpha_{1} \beta_{1} \ldots \alpha_{n} \beta_{n}}^{\mu_{1} \nu_{1} \ldots \mu_{n} \nu_{n}} \prod_{r=1}^{n} R_{\mu_{r} \nu_{r}}^{\alpha_{r} \beta_{r}}\,,
\end{align}
and study the generalized covariant entropy bound, following the proposal in Sec.~\ref{sec:beyondGR}.

\subsection{Wald-Jacobson-Myers entropy and covariant entropy bound}

The generalized covariant entropy bound proposed in Sec.~\ref{sec:beyondGR} for theories beyond GR assumes a well-defined formula of black hole entropy that can be applied to not only black hole horizons but also a family of connected $(D-2)$-dimensional spatial surfaces that generates a generalized lightsheet. In Sec.~\ref{sec:examples} we adopted the Wald formula (\ref{eqn:Waldentropy}) for $f(R)$ gravity and a canonical scalar-tensor theory.

Unfortunately, in the presence of Lovelock terms Wald entropy is known to be ambiguous for non-stationary black hole horizons~\cite{Jacobson:1993vj}. In $D$ dimensions the $n$-th Lovelock term $\sqrt{-g}\mathcal{L}_{n}[g_{\mu\nu}]$ with $n>(D-1)/2$ is total derivative and thus does not contribute to the equation of motion. Nonetheless, by explicit calculation one can see that such Lovelock terms with $n>(D-1)/2$ can contribute to Wald entropy. For example, in $4$ dimensions the second Lovelock term is total derivative but its contribution to Wald entropy is
\begin{align}
 R^{pq}_{\ \ pq} =  R^{(2)} + \left( \mbox{terms bilinear in } K_{1} \mbox{ and } K_{2} \right)\,, \label{eqn:wald-2ndlovelock-4d}
\end{align}
where $R^{\mu\nu}_{\ \ \rho\sigma}$ is the $4$-dimensional Riemann curvature, $R^{(2)}$ is the Ricci scalar of the induced metric on the $2$-dimensional horizon cross section with coordinates labeled by $p$ and $q$, and $K_{1,2}$ are second fundamental forms along two null directions orthogonal to the horizon cross section~\footnote{See e.g. \cite{Gourgoulhon:2005ng} for the double-null decomposition of the Riemann curvature.}. The integral of $R^{(2)}$ over the $2$-dimensional horizon cross section is constant and thus its variation vanishes. On the other hand, the terms bilinear in $K_{1}$ and $K_{2}$ in (\ref{eqn:wald-2ndlovelock-4d}) lead to a non-vanishing and non-constant contribution to Wald entropy. Considering the fact that the second Lovelock term with an arbitrary constant coefficient does not contribute to the equation of motion in $4$ dimensions, this result shows that equivalent actions give different Wald entropies for non-stationary black hole horizons. On the other hand, for stationary black hole horizons the $K_{1}K_{2}$ terms vanish~\footnote{For a stationary Killing horizon, one of the two null directions orthogonal to the horizon cross section is along the horizon and thus either $K_{1}$ or $K_{2}$ vanishes.} and thus Wald entropy is unique up to a constant. In the presence of Lovelock terms it is thus concluded that the expression of Wald entropy (\ref{eqn:Waldentropy}) can be used only for stationary black hole horizons. For more general situations such as non-stationary black hole horizons and general connected $(D-2)$-dimensional spatial surfaces, one thus needs to correct Wald formula.

In the present paper we adopt the following formula~\cite{Jacobson:1993xs} of black hole entropy in Lovelock gravity: 
\begin{align}
S_{\rm bh}(B) =  \int_{B} d^{D-2}x \sqrt{h} s_{\rm bh}\,,
\end{align}
where $B$ is a connected $(D-2)$-dimensional spatial surface, $h$ is the determinant of the induced metric $h_{pq}$ on $B$, $s_{\rm bh}$ is given by 
\begin{equation}
 s_{\rm bh} = \frac{1}{4}\sum_{m=1}^{[(D-1)/2]}m\alpha_{m} \mathcal{L}_{m-1}[h_{pq}]\,, \label{eqn:sbh-JM}
\end{equation}
$\mathcal{L}_{m-1}[h_{pq}]$ is the $(m-1)$-th Lovelock term in $(D-2)$ dimensions applied to $h_{pq}$ and $\mathcal{L}_{0}[h_{pq}]=1$. This form of black hole entropy, when $B$ is set to be a horizon cross section, is called Jacobson-Myers (JM) entropy. JM entropy coincides with Wald entropy when $B$ is chosen to be a stationary black hole horizon. For non-stationary linear perturbations around a stationary black hole spacetime, JM entropy of the perturbed black hole horizon has the following two properties suggesting that JM entropy qualifies as entropy of not only stationary but also non-stationary black holes in Lovelock gravity. One is that JM entropy does not decrease along a future pointing generator of an event horizon~\cite{Kolekar:2012tq}, meaning that the analogue of the classical second law of black holes holds for JM entropy. The other is that the generalized second law holds for JM entropy under a set of reasonable assumptions~\cite{Sarkar:2013swa}. For these reasons it is expected that JM entropy serves as black hole entropy in Lovelock gravity even for non-stationary black holes. We thus adopt (\ref{eqn:sbh-JM}) to formulate the generalized covariant entropy bound in Lovelock gravity.

\subsection{EGB gravity with spherical symmetry}

In the rest of this section, for simplicity we consider Einstein-Gauss-Bonnet (EGB) gravity, for which  the gravitational action is given by the first two Lovelock terms as 
\begin{equation}
 I = \frac{1}{16\pi}\int d^Dx\sqrt{-g}\left( R + \alpha \mathcal{L}_{2}[g_{\mu\nu}] \right)\,,
\end{equation}
where $\mathcal{L}_{2}[g_{\mu\nu}] = R^2-4R^{\mu\nu}R_{\mu\nu}+R^{\mu\nu\rho\sigma} R_{\mu\nu\rho\sigma}$ and we have set $\alpha_1 = 1$ and $\alpha_2 = \alpha$. Hence $s_{\rm bh}$ for JM entropy given in (\ref{eqn:sbh-JM}) is reduced to 
\begin{equation}
 s_{\rm bh} = \frac{1}{4}\left[ 1 + 2\alpha R^{(D-2)}\right]\,, \label{eqn:sbh-JM-egb}
\end{equation}
where $R^{(D-2)}$ is the Ricci scalar of $h_{pq}$ on $B$. We assume 
\begin{equation}
 \alpha > 0\,,  \label{eqn:alpha-positive}
\end{equation}
motivated by string theory (see e.g. \cite{Gross:1986iv,Zwiebach:1985uq}). This assumption implies that black hole entropy increases by the effect of higher derivative terms for a fixed horizon area. If we consider black hole entropy as the logarithm of the number of black hole states, the increase of black hole entropy due to higher derivative terms that encapsulate extra heavy modes is rather natural.

For further simplicity we restrict our consideration to spherically symmetric configurations. Introducing double null coordinates the metric is in general written as 
\begin{equation}
 ds^2 = -2 e^{-f(u,v)}dudv + r^2(u,v)\Omega_{pq}dx^pdx^q\,, \label{eqn:metric-spherical-doublenull}
\end{equation}
where $\Omega_{pq}dx^pdx^q$ is the metric of the unit $(D-2)$-sphere. The JM formula (\ref{eqn:sbh-JM-egb}) is then reduced to 
\begin{equation}
 s_{\rm bh} = \frac{1}{4}\left[ 1 + \frac{2(D-2)(D-3)}{r^2}\right]\,, \label{eqn:sbh-JM-egb-spherical}
\end{equation}
The generalized Misner-Sharp mass~\cite{Maeda:2006pm} is defined by 
\begin{align}
 m & = \frac{(D-2)\mathrm{a}_{D-2}}{16\pi}\left[r^{D-3}(1+2e^{f}r_{,u}r_{,v}) \right. \nonumber\\
& \qquad \qquad \quad \left. + \tilde{\alpha} r^{D-5}(1+2e^{f}r_{,u}r_{,v})^2\right]\,,
\end{align}
where $\mathrm{a}_{D-2}=(D-1)\pi^{\frac{D-1}{2}}/\Gamma((D+1)/2)$ is the surface area of the unit $(D-2)$-sphere and we have introduced $\tilde{\alpha}=(D-3)(D-4)\alpha$ for brevity of some expressions. It follows that $m$ is non-negative on a spacelike or lightlike hypersurface whose intersection with the center ($r=0$) is regular, provided that the dominant energy condition holds~\cite{Maeda:2007uu}. Furthermore, the above definition of $m$ can be rewritten as
\begin{equation}
 1 + \frac{2\tilde{\alpha}}{r^2}(1+2e^{f}r_{,u}r_{,v}) = \pm \sqrt{1 + \frac{64\pi\tilde{\alpha}m}{(D-2)\mathrm{a}_{D-2}r^{D-1}}}\,, \label{eqn:egb-spherical-twobranches}
\end{equation}
and this equation defines two branches of spherically symmetric solutions of the theory: the GR branch for the ``$+$'' sign and the non-GR branch for the ``$-$'' sign. The two branches are distinct in the sense that they can merge only at a curvature singularity, provided that the null energy condition is strictly satisfied for radial null vectors~\cite{Nozawa:2007vq}.

Since it is the GR branch that reduces to GR in the $\alpha\to 0$ limit, we shall hereafter restrict our consideration to the GR branch. In the GR branch the null energy condition for radial null vectors implies the null convergence condition for radial null vectors~\cite{Nozawa:2007vq} (see (\ref{eqn:8piTuu-egb-spherical}) below). This in particular means that for a black hole spacetime in the GR branch, an apparent horizon is always on or inside the event horizon, provided that the null energy condition holds.

\subsection{Proof in EGB gravity with spherical symmetry}
\label{subsec:proof-egb}

In this subsection we shall consider the GR branch of EGB gravity with spherical symmetry and prove the generalized covariant entropy bound proposed in Sec.~\ref{sec:beyondGR} by employing the formula (\ref{eqn:sbh-JM-egb}) or (\ref{eqn:sbh-JM-egb-spherical}) for $s_{\rm bh}$. For the proof, we employ the coordinate system $(x,\lambda)$ on a generalized lightsheet $L$ as introduced in Sec.~\ref{sec:beyondGR}, and make the following two assumptions on $L$. 
\begin{align}
 \text{(i-gb)} & & & \frac{\partial s}{\partial \lambda}(x,\lambda)  \leq 2\pi \mathcal{T}(x,\lambda)\,, \nonumber\\
 \text{(ii-gb)} & & & s(x,0) \leq - s_{\mathrm{bh}}(x,0)\Theta(x,0)\,, \label{eqn:assumptions-EGB}
\end{align}
where $\mathcal{T}=T_{ab}k^ak^b$, $T_{ab}$ is the stress-energy tensor of matter fields and $\Theta$ is defined by (\ref{eqn:expansiongeneralized}) with (\ref{eqn:sbh-JM-egb}) or (\ref{eqn:sbh-JM-egb-spherical}). These two assumptions correspond to (\ref{eqn:assumptions-GR}) in GR, (\ref{eqn:assumptions-f(R)-originalframe-i}) and (\ref{eqn:assumptions-f(R)-originalframe-ii}) in $f(R)$ gravity, and (\ref{eqn:assumptions-ST-originalframe-i}) and (\ref{eqn:assumptions-ST-originalframe-ii}) in canonical scalar-tensor theory.

As we assumed the spherical symmetry, we adopt the metric (\ref{eqn:metric-spherical-doublenull}). Furthermore, we redefine the null coordinate $u$ so that $u=\lambda$ on $L$, where $\lambda$ is the normalized affine parameter introduced in Sec.~\ref{sec:beyondGR}. This in particular implies that $L$ is described by $v=v_{L}$ and $0\leq u \leq 1$, where $v_{L}$ is a constant, and that $f_{,u}=0$ on $L$. On the other hand, a priori there is no relation between $u$ and $\lambda$ away from $L$ and thus e.g. $f_{,uv}$ does not vanish on $L$ in general. A key equation is
\begin{equation}
 8\pi T_{uu} = R_{uu} \left[ 1 + \frac{2\tilde{\alpha}}{r^2}(1+2e^{f}r_{,u}r_{,v}) \right]\,,
  \label{eqn:8piTuu-egb-spherical}
\end{equation}
where
\begin{equation}
 R_{uu} = -\frac{D-2}{r}(r_{,uu}+f_{,u}r_{,u})\,.
\end{equation}
In the GR branch, i.e. for the ``$+$'' sign in (\ref{eqn:egb-spherical-twobranches}), (\ref{eqn:8piTuu-egb-spherical}) and $T_{uu}\geq 0$ implies $R_{uu}\geq 0$. Namely, the null energy condition for radial null vectors implies the null convergence condition for radial null vectors, as already mentioned in the previous subsection. Restricting to the generalized lightsheet $L$, we thus have
\begin{equation}
 \mathcal{T} = -\frac{D-2}{8\pi}\frac{r_{,uu}}{r}\left[ 1 + \frac{2\tilde{\alpha}}{r^2}(1+2e^{f}r_{,u}r_{,v}) \right]\ \ \mbox{on L}\,, \label{eqn:calT-EGB-spherical}
\end{equation}
and $r_{,uu}\leq 0$ on $L$.

We now proceed to the proof. First, by integrating the assumption $\text{(i-gb)}$ in (\ref{eqn:assumptions-EGB}) once and using (\ref{eqn:calT-EGB-spherical}), one obtains
\begin{widetext}
\begin{equation}
 s(\lambda) - s(0) \leq -\frac{D-2}{4}\int_{0}^{\lambda}du \frac{r_{,uu}}{r}\left[ 1 + \frac{2\tilde{\alpha}}{r^2}(1+2e^{f}r_{,u}r_{,v}) \right] 
  = -\frac{D-2}{4}\left[\frac{r_{,u}}{r}\left( 1 + \frac{2\tilde{\alpha}}{r^2} \right)\right]^{u=\lambda}_{u=0}  - F(\lambda)\,, \label{eqn:proof-egb-inequality1}
\end{equation}
where $v=v_{L}$ is imposed and 
\begin{equation}
 F(\lambda) = \left.\frac{D-2}{4}\int_{0}^{\lambda}du
  \left[\left(\frac{r_{,u}}{r}\right)^2\left( 1 + \frac{6\tilde{\alpha}}{r^2}\right)
 + 4\tilde{\alpha}\frac{r_{,uu}}{r^3}e^{f}r_{,u}r_{,v}\right]\right|_{v=v_{L}}\,. \label{eqn:def-F(lambda)}
\end{equation}
\end{widetext}
Second, the assumption $\text{(ii-gb)}$ in (\ref{eqn:assumptions-EGB}) is written as
\begin{equation}
 s(0) \leq \left.-\frac{D-2}{4}\frac{r_{,u}}{r}\left( 1 + \frac{2\tilde{\alpha}}{r^2} \right)\right|_{u=0,v=v_{L}}\,. \label{eqn:proof-egb-inequality2}
\end{equation}
By combining (\ref{eqn:proof-egb-inequality1}) and (\ref{eqn:proof-egb-inequality2}), one obtains
\begin{equation}
 s(\lambda) \leq \left. -\frac{D-2}{4}\frac{r_{,u}}{r}\left( 1 + \frac{2\tilde{\alpha}}{r^2} \right)\right|_{u=\lambda,v=v_{L}} - F(\lambda)\,. 
\end{equation}
Multiplying this inequality by $\mathrm{a}_{D-2} r^{D-2}$ and then integrating it from $\lambda=0$ to $\lambda=1$, where $\mathrm{a}_{D-2}$ is the surface area of the unit $(D-2)$-sphere, one obtains
\begin{equation}
 S_{L} \leq S_{\rm bh}(B) - S_{\rm bh}(B') - \mathrm{a}_{D-2}\int_{0}^{1}d\lambda F(\lambda) r^{D-2}(\lambda)\,.
\end{equation}
Provided that 
\begin{equation}
 \int_{0}^{1}d\lambda F(\lambda) r^{D-2}(\lambda) \geq 0\,,  \label{eqn:additionalassumption-EGB}
\end{equation}
this completes the proof.

Finally, let us now argue that (\ref{eqn:additionalassumption-EGB}) is a reasonable assumption. Indeed, whenever the GB correction to GR is subdominant, the leading term in $F(\lambda)$ defined by (\ref{eqn:def-F(lambda)}) is $F(\lambda) \simeq \int_{0}^{\lambda}du (D-2)(r_{,u}/r)^2/4|_{v=v_L} \geq 0$, which leads to (\ref{eqn:additionalassumption-EGB}). Moreover, even when the GB correction is not subdominant, one can show that (\ref{eqn:additionalassumption-EGB}) holds if the lightsheet $L$ does not enter a trapped region. In this case we have $e^fr_{,u}r_{,v}\leq 0$ on $L$. Combining this with (\ref{eqn:alpha-positive}) and $r_{,uu}\leq 0$ on $L$, which in the GR branch is implied by the null energy condition, it is concluded that $F(\lambda)\geq 0$ on $L$ and that (\ref{eqn:additionalassumption-EGB}) holds.

\subsection{Corollary: generalized second law in EGB gravity with spherical symmetry}

In this subsection we apply the generalized covariant entropy bound proved in the previous subsection to a black hole spacetime in the GR branch of EGB gravity with spherical symmetry to prove the generalized second law.

Before the proof of the generalized second law, let us prove the classical second law of black hole event horizons for JM entropy under the null energy condition. As mentioned in the previous subsection, for a black hole spacetime in the GR branch, an apparent horizon is always on or inside the event horizon, provided that the null energy condition holds for radial null vectors. This means that the event horizon is either outside or on the outermost apparent horizon. Hence, the area of the event horizon does not decrease in time and $e^fr_{,u}r_{,v}\leq 0$ on the event horizon. Since 
\begin{equation}
 \frac{d}{dr}(s_{\rm bh} r^{D-2}) = \frac{D-2}{4}r^{D-3}\left(1+\frac{2\tilde{\alpha}}{r^2}\right)\,,
\end{equation}
the definition of GR branch, i.e. (\ref{eqn:egb-spherical-twobranches}) with the ``$+$'' sign, the assumption (\ref{eqn:alpha-positive}) and the inequality $e^fr_{,u}r_{,v}\leq 0$ shown above imply that $s_{\rm bh} r^{D-2}$ is an increasing function of $r$ and that not only the horizon area but also $S_{\rm bh}$ are non-decreasing along the event horizon towards the future. Namely, the classical second law holds for JM entropy of black hole event horizons, provided that the GR branch is chosen and that the null energy condition holds.

Let us now prove the generalized second law, which is stronger than the classical second law. We assume the null energy condition and the dominant energy condition. As shown above, $e^fr_{,u}r_{,v}\leq 0$ on a black hole event horizon and JM entropy evaluated on the event horizon does not decrease towards the future. Therefore, one can choose a part of the event horizon as a past-directed generalized lightsheet $L$ and the generalized covariant entropy bound holds for $L$ (see the last paragraph of the previous subsection). Let $B$ and $B'$, respectively, be the future and past boundaries of $L$. The generalized covariant entropy bound then implies that 
\begin{equation}
 S_{L} \leq S_{\rm bh}(B) - S_{\rm bh}(B')\,.
\end{equation}
Under the dominant energy condition and the second law of thermodynamics applied to the region outside the black hole, we have 
\begin{equation}
 S_{\rm matter}(B) + S_{L} \geq S_{\rm matter}(B')\,,
\end{equation}
where $S_{\rm matter}(B)$ and $S_{\rm matter}(B')$ are the entropies of matter outside the event horizon on spacelike hypersurfaces intersecting with $B$ and $B'$, respectively, and we suppose that these two hypersurfaces do not intersect. Combining the two inequality, we obtain
\begin{equation}
 S_{\rm bh}(B) + S_{\rm matter}(B) \geq S_{\rm bh}(B') + S_{\rm matter}(B')\,.
\end{equation}
This is the generalized second law.

\section{Summary and discussion}
\label{sec:summary}

In this paper we have extended the covariant entropy bound that was originally formulated in general relativity (GR) to gravitational theories beyond GR, as prescribed in Sec.~\ref{sec:beyondGR} in full generality. As concrete and simple examples, we have constructed the generalized covariant entropy bound in $f(R)$ gravity and a canonical scalar-tensor theory, by using Wald entropy instead of Bekenstein-Hawking entropy, and provided a proof in a thermodynamic limit in each theory. As a more non-trivial example, we have considered Lovelock gravity, for which Wald entropy turned out to be ambiguous in non-stationary spacetimes. We have therefore employed Jacobson-Myers (JM) entropy, which can be applied to non-stationary spacetimes and is known to satisfy the classical second law and the generalized second law up to linear order in non-stationary perturbations around a stationary black hole background. With JM entropy, we have proposed the generalized covariant entropy bound in Lovelock gravity and then provided a proof for spherically symmetric configurations of Einstein-Gauss-Bonnet (EGB) gravity, which is the simplest non-trivial subset of Lovelock gravity, under a set of reasonable assumptions. The assumptions of the proof are summarized in (\ref{eqn:assumptions-EGB}) and (\ref{eqn:additionalassumption-EGB}). The ones in (\ref{eqn:assumptions-EGB}) are direct analogues of those in general relativity (see (\ref{eqn:assumptions-GR}) and justifications of them in subsections \ref{subsec:assumption-i} and \ref{subsec:assumption-ii}) and we have also argued that (\ref{eqn:additionalassumption-EGB}) is a reasonable assumption (see the last paragraph of subsection~\ref{subsec:proof-egb}). As a corollary of the proof, we have also shown that JM entropy satisfies the generalized second law in the GR branch of EBG gravity with spherical symmetry at the fully nonlinear order, i.e. without the assumption of small non-stationarity.

These results serve as supporting evidence for our proposal of the generalized covariant entropy bound to some extent. For Lovelock gravity, however, we have only considered EGB gravity with spherical symmetry. Also, we have only discussed a thermodynamic limit and a fluid-like matter, whose entropy flow can be represented by a timelike vector. If we think of the covariant entropy bound as a holographic property of gravity, then the bound is expected to hold in a more general setting of matter. One of the future works is therefore to remove these restrictions. It is certainly interesting to extend the bound to other theories of gravity as well. Another important future work is to find the structural reason why the generalized covariant entropy bound holds for various theories of gravity, and to identify the space of theories where the bound holds.

\section*{acknowledgement}
The work of S.M. was supported by Japan Society for the Promotion of Science Grants-in-Aid for Scientific Research No.~17H02890, No.~17H06359, and by World Premier International Research Center Initiative, MEXT, Japan.

\end{document}